\documentclass{JHEP3}

\usepackage{amsmath}
\usepackage{amsfonts}
\usepackage{amssymb}
\usepackage{graphicx}

\DeclareMathOperator{\sign}{sign}
\newcommand{\nvec}[1]{\boldsymbol{#1}}
\newcommand{\abs}[1]{\left\vert #1 \right\vert}
\newcommand{\ud}{\mathrm{d}}
\newcommand{\rt}{r}
\newcommand{\vrt}{\nvec{r}}
\newcommand{\dE}{\Delta E}
\newcommand{\kt}{k_T}
\newcommand{\vkt}{\nvec{k}_T}
\newcommand{\vktbar}{\bar{\nvec{k}}_T}
\newcommand{\dEa}{\Delta E}
\newcommand{\dEb}{\Delta\bar{E}}
\newcommand{\dEp}{\Delta E_+}
\newcommand{\dEm}{\Delta E_-}
\newcommand{\ka}{k}
\newcommand{\kb}{\bar{k}}
\newcommand{\kat}{k_T}
\newcommand{\kbt}{\bar{k}_T}
\newcommand{\vkat}{\nvec{k}_T}
\newcommand{\vkbt}{\nvec{\bar{k}}_T}
\newcommand{\etaa}{\eta}
\newcommand{\etab}{\bar{\eta}}
\newcommand{\taua}{\tau}
\newcommand{\taub}{\bar{\tau}}
\newcommand{\sigmadip}{\hat{\sigma}}
\newcommand{\sigmatot}{\sigma}

\preprint{HD-THEP-09-13\\BI-TP 2009/17\\ZU-TH-12-09}
\title{Ioffe Times in DIS from a Dipole Model Fit}
\author{Carlo Ewerz\,$^{a,b,c,1}$, Andreas von Manteuffel\,$^{d,2}$, Otto Nachtmann\,$^{a,3}$
\\
$^{a}$ 
Institut f\"ur Theoretische Physik, Universit\"at Heidelberg,\\
\phantom{$^b$} Philosophenweg 16, D-69120 Heidelberg, Germany\\
$^{b}$
ExtreMe Matter Institute EMMI, GSI Helmholtzzentrum f\"ur Schwerionenforschung,\\
\phantom{$^a$} 
Planckstra{\ss}e 1, D-64291 Darmstadt, Germany\\
$^c$
Fakult\"at f\"ur Physik, Universit\"at Bielefeld, 
D-33615 Bielefeld, Germany\\
$^d$
Institut f\"ur Theoretische Physik, Universit\"at Z\"urich, \\
\phantom{$^d$}
Winterthurerstr.\ 190, CH-8057 Z\"urich, Switzerland
\\
$^1$E-mail: \email{C.Ewerz@thphys.uni-heidelberg.de}\\
$^2$E-mail: \email{manteuffel@physik.uzh.ch}\\
$^3$E-mail: \email{O.Nachtmann@thphys.uni-heidelberg.de}
}

\abstract{
We present a study of Ioffe times in deep inelastic electron-proton 
scattering. We deduce `experimental' Ioffe-time distributions 
from the small-$x$ HERA data as described by a particular 
colour-dipole-model fit. We show distributions for three representative
$\gamma^* p$ c.\,m.\ energies $W$ and various values of 
the photon virtuality $Q^2$. These distributions are rather broad for 
transversely and very narrow for longitudinally polarised 
virtual photons. The Ioffe times for $W=150~\mbox{GeV}$, 
for example, range from around $10^3$ fm for 
$Q^2= 1~\mbox{GeV}^2$ to around 10~fm for 
$Q^2 = 100~\mbox{GeV}^2$. 
Based on our results we discuss consequences for the 
limitations of applicability of the dipole picture. 
}


\begin{document}

\section{Introduction}
\label{sec:intro}

In this article we shall present a quantitative study of the space-time 
structure of deep inelastic electron- and positron-proton scattering (DIS) 
\begin{equation}\label{processep}
e^\pm + p  \rightarrow e^\pm + X
\end{equation}
as measured extensively in HERA experiments 
\cite{Breitweg:2000yn,Adloff:2000qk,Chekanov:2001qu,Adloff:2003uh,Chekanov:2003yv}. 
We shall be concerned with the kinematic region where only exchange 
of a virtual photon between the leptons and the hadrons matters. Thus, 
as usual, with the reaction \eqref{processep} we study in fact the absorption 
of a virtual photon on the proton 
\begin{equation}\label{processapintro}
\gamma^\ast + p \rightarrow X  \,.
\end{equation}
The total cross section for \eqref{processapintro} is, apart from 
kinematic factors, given by the imaginary part of the amplitude 
for forward Compton scattering of a virtual photon on the proton, 
\begin{equation}
\label{gampgamp}
\gamma^* + p \rightarrow \gamma^* + p \,.
\end{equation}

The study of the space-time structure of this process has a long 
history going back to the discussions of the vector-meson-dominance 
(VMD) model in the 1960s and in particular to articles by Gribov et al.\ 
\cite{Gribov:1965hf} and Ioffe \cite{Ioffe:1969kf}. In \cite{Ioffe:1969kf} 
the question was posed where and when the initial virtual photon 
is absorbed, that is, where it fluctuates into `hadronic stuff' 
and where the `hadronic stuff' fluctuates back to give 
the final virtual photon. In this article we shall give 
quantitative answers to this question, as far as the region of small $x$ 
is concerned, using the HERA data. 
In fact we shall use a fit to these data in a particular colour dipole 
model. 

Let us first make some more historical remarks. It is an old idea 
that the scattering of a highly-energetic photon on a hadron may essentially 
be considered as a strong interaction process where the photon 
acts in some way as a hadron. 
Vector mesons as intermediary particles in the coupling of photons 
to nucleons were introduced in 
\cite{Nambu1957,Frazer:1959gy,Frazer:1960zzb} in order to understand 
properties of the electromagnetic form factors. An important role was 
assigned to vector mesons in strong interactions in \cite{Sakurai:1960ju}. 
Then in \cite{GellMann:1961tg} the vector-meson-dominance 
model was proposed. There it is assumed that whenever a photon 
couples to hadrons it first converts to the vector mesons $\rho$, $\omega$, 
$\phi$ (proposed / known at the time) with universal coupling constants. 
These vector mesons have then normal hadronic reactions. 
The VMD model was rather successful in describing many results from 
photon-hadron reactions. For reviews see, for instance, 
\cite{Sakurai-book,Bauer:1977iq}. But the results from deep inelastic 
scattering presented problems. Indeed, in \cite{Sakurai:1969ss} a simple form of 
this VMD picture was applied to DIS and predictions were made which, 
however, were subsequently disproven by experiments. 
This led to the formulation of the concept of generalized vector meson 
dominance, see \cite{Schildknecht:1972ej,Bauer:1977iq,Schildknecht:2005xr} 
and references therein. 
The key idea in these descriptions is that the photon fluctuates into 
a series of vector mesons which subsequently scatter off the proton. 
Quite obviously, the lifetime of the fluctuation of the photon into a vector 
meson needs to be sufficiently long for that picture to be consistent. 

The colour dipole model \cite{Nikolaev:1990ja,Nikolaev:et,Mueller:1993rr} 
builds on similar ideas but is motivated to a large 
extent by perturbative QCD. In this picture the reaction \eqref{processapintro} 
is viewed as a two-step process. In the first step the photon splits into a 
quark-antiquark pair which represents the colour dipole. 
Subsequently that pair scatters on the proton, this second step being a purely
hadronic reaction. For the applicability of the dipole model it is crucial 
that the lifetime of the fluctuation of the photon into the colour dipole 
is much larger than the typical timescale of the dipole-proton interaction. 
In the context of the dipole picture the lifetime of the dipole fluctuation is 
usually referred to as the Ioffe time. 
We may note in parentheses, however, that Ioffe in his original paper 
\cite{Ioffe:1969kf} was concerned with a somewhat different time as we shall 
comment on below. It is the aim of the present paper to 
study the distribution of Ioffe times and their dependence on the 
kinematical parameters of the photon-proton scattering process 
using the HERA data on DIS. 

The splitting of a photon of high enough virtuality $Q^2$ 
into a quark-antiquark pair can be calculated in perturbation theory. 
The subsequent scattering of the 
colour dipole off the proton, on the other hand, 
is a genuinely nonperturbative process. Therefore, this second step 
of the $\gamma^* p$ scattering process 
needs to be described by suitable models. A variety of such models has been 
constructed, see \cite{GolecBiernat:1998js,Bartels:2002cj,Iancu:2003ge} 
for some prominent examples and \cite{Donnachie:2002en,Motyka:2008jk} 
for overviews. These models are very successful in describing the 
structure functions measured at HERA. 

But despite the impressive phenomenological success of dipole 
models there are some caveats. In \cite{Ewerz:2004vf,Ewerz:2006vd} the 
foundations of the dipole picture were examined in detail. It was shown that 
a number of assumptions and approximations is required to arrive at the 
dipole picture. These results naturally raise the question about the range 
of validity of these approximations and assumptions. 
In \cite{Ewerz:2006an,Ewerz:2007md} it was found that already the general 
formulae of the standard 
dipole approach allow one to derive stringent bounds on various ratios of structure 
functions. These bounds were then used to determine the kinematic region where 
the dipole picture is compatible with the data. 
An important point in this connection concerns the variables on which the 
dipole-proton cross section depends. The natural variables were found to 
be $r$, the transverse size of the dipole, and $W$, the c.\,m.\ energy of 
the dipole-proton scattering. Using this functional dependence for the 
dipole-proton cross section we obtained the following result: 
For $\gamma^\ast p$ c.\,m.\ energies $W$ in the 
range of 60 to 240~GeV the standard dipole picture fails to be compatible with the 
HERA data for photon virtualities $Q^2$ larger than about 
$100$ to $200$~GeV$^2$ \cite{Ewerz:2007md}. Clearly, these large virtualities 
correspond to relatively short Ioffe times. In the present paper we want to 
study this relation between Ioffe times and the kinematical parameters 
in more detail by calculating the distribution of Ioffe times and the 
corresponding contributions to the structure functions for given 
values of $Q^2$ and $W$. 

Let us, therefore, consider the reaction \eqref{processapintro} at high c.\,m.\ energy 
in the proton rest system.
The virtual photon $\gamma^\ast$ of $4$-momentum $q=(q^0,\nvec{q})$ fluctuates
into a quark and antiquark of $4$-momenta $k=(k^0,\nvec{k})$ and
$k'=(k'^0,\nvec{k}')$ respectively, where $3$-momentum is conserved, 
$\nvec{q}=\nvec{k}+\nvec{k}'$. Energy, however, is not conserved. 
Instead, there is an energy mismatch
\begin{equation}\label{dEintro}
\dE = k^0 + k'^0 - q^0
\end{equation}
between the quark-antiquark pair and the virtual photon. 
According to the uncertainty relation such a 
fluctuation cannot live longer than the time
\begin{equation}\label{tauintro}
\tau = \frac{1}{\dE} = \frac{1}{k^0+k'^0-q^0}\,.
\end{equation}
In the following we shall call $\tau$ the Ioffe time for the 
initial $\gamma^*$. 
Discussions of Ioffe time distributions in the context of the 
operator product expansion can be found in \cite{Braun:1994jq}. 
In \cite{Kovchegov:2001dh} the Ioffe-time structure of the gluon 
distribution function in the double logarithmic approximation 
was considered. 

In applications of the dipole model one frequently finds simple estimates for
Ioffe times, and if they turn out to be of the order of several femtometers or larger
this is taken as justification for using the dipole model. But the actual situation 
is more involved. 
We shall find that even for a fixed kinematical point for reaction \eqref{processapintro}
one has a whole distribution of Ioffe times.
In fact, the $\gamma^\ast p$ total absorption cross section is most conveniently 
obtained from the imaginary part of the forward scattering amplitude 
for the reaction \eqref{gampgamp}, 
$\gamma^\ast p \to \gamma^\ast p$. As a consequence, we shall have to 
deal with {\em two} Ioffe times, one for the initial state
$\gamma^\ast$ and one for the final state $\gamma^\ast$.
Both times have distributions which also depend on the polarisation,
transverse or longitudinal, of the $\gamma^\ast$.
We shall calculate such distributions in the following from the HERA data 
using a phenomenological dipole model which describes the data quite well. 
We shall choose the model of Golec-Biernat and W\"usthoff (GBW) 
\cite{GolecBiernat:1998js} for this purpose. 

In section \ref{sec:picture} we review the relevant formulae of the 
dipole picture and define the $\dE$ and Ioffe time $\tau$ distributions 
in this framework. In section \ref{sec:results} we recall the basics of 
the GBW model for the dipole cross section. We then present numerical 
results for the $\Delta E$ and the corresponding Ioffe-time distributions. 
Our conclusions are drawn in section \ref{sec:conclusions}. 
In two appendices we present the details of our calculations. 

\boldmath
\section{Ioffe time and $\Delta E$ distributions in the dipole picture}
\unboldmath
\label{sec:picture}

We use the standard formulae for the kinematics and for the definitions of
structure functions of deep inelastic electron- and positron-proton scattering 
\eqref{processep}, see for instance \cite{Nachtmann:1990ta}.
For momentum transfers squared $Q^2 \lesssim 1000~\text{GeV}^2$ it is sufficient
to consider only the exchange of a virtual photon between the lepton and the proton in
\eqref{processep}.
Thus, the reaction which we shall study in the following is the absorption of a
virtual photon $\gamma^\ast$ on the proton,
\begin{equation}\label{processap}
\gamma^\ast(q) + p(p) \rightarrow X(p')\,,
\end{equation}
where we indicate the $4$-momenta in brackets.
The c.m.\ energy for this reaction is denoted by $W$, the virtuality of the 
$\gamma^\ast$ by $Q^2$. For these and for Bjorken's scaling variable $x$ 
we have 
\begin{equation}
\label{defkin}
\begin{split}
W^2 &= (p+q)^2\,, \\
Q^2 &= -q^2\,, \\
x &= \frac{Q^2}{2 p \cdot q} = \frac{Q^2}{W^2+Q^2-m_p^2}\,.
\end{split}
\end{equation}
The proton in \eqref{processap} is supposed to be unpolarised, but the virtual photon
can have transverse or longitudinal polarisation.
The corresponding total cross sections are $\sigma_T(W,Q^2)$ and
$\sigma_L(W,Q^2)$, respectively.
The structure function $F_2$ is
\begin{equation}\label{f2diphand}
F_2(W,Q^2) = \frac{Q^2}{4 \pi^2 \alpha_{\rm em}}
   \left[ \sigma_T(W,Q^2) + \sigma_L(W,Q^2) \right] (1-x)
+ \mathcal{O}\left(\frac{m_p^2}{W^2}\right) \,.
\end{equation}
For small Bjorken-$x$, $x\ll 1$, this simplifies to
\begin{equation}\label{f2dipsimple}
F_2(W,Q^2) = \frac{Q^2}{4\pi^2\alpha_{\mathrm{em}}}
  \left[ \sigma_T(W,Q^2) + \sigma_L(W,Q^2) \right]\,.
\end{equation}
In the following we shall use this simpler relation since we shall only consider
data for $x\ll 1$.
In order to obtain the standard dipole model for the cross sections $\sigma_{T,L}$ we
can relate them first to the imaginary part of the virtual Compton 
forward scattering amplitude, 
\begin{equation}
\label{Comptonmitarg}
\gamma^* (q) + p(p) \rightarrow \gamma^*(q) + p(p) \,.
\end{equation}
The latter is then represented as the initial $\gamma^\ast$ 
splitting into a $q\bar q$
pair, this pair scattering on the proton, and the $q\bar q$ subsequently fusing
into the final state $\gamma^\ast$, see figure \ref{fig:dipolediag}. 
Note that this figure is to be read from right to left in order to be in
complete analogy with the occurrence of the various factors in the amplitudes 
in the equations below.
\FIGURE{
\includegraphics[width=0.7\textwidth]{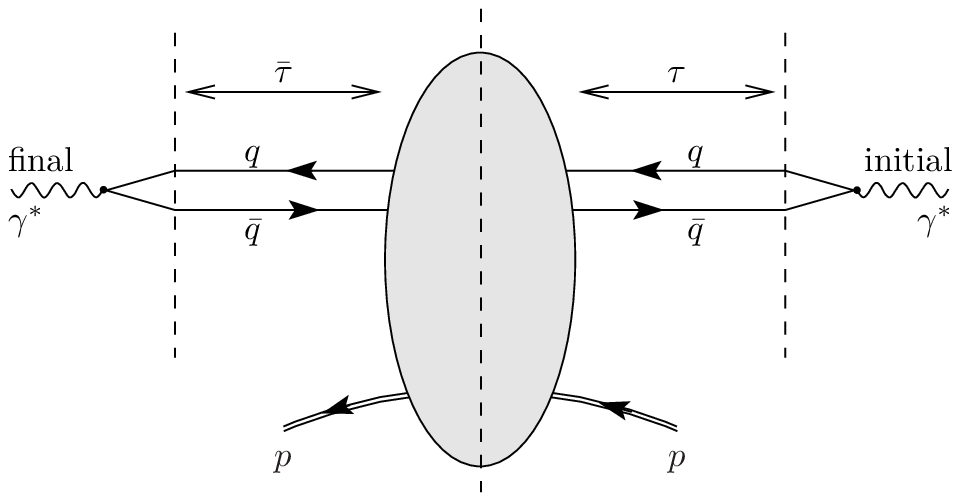}
\caption{Basic diagram for the description of the cross sections $\sigma_{T,L}$ of
$\gamma^\ast p$ scattering in the standard dipole approach 
\label{fig:dipolediag}
}}
With the assumptions spelled out in detail in section 6 of \cite{Ewerz:2006vd} the
diagram of figure \ref{fig:dipolediag} gives 
\begin{align}
\label{xstnoq2}
\sigma_T(W,Q^2) &= \sum_q \int \ud^2 r \int_0^1\ud\alpha \sum_{\lambda,\lambda'}
  \left(\psi_{\lambda\lambda'}^{(q)\,\pm}(\alpha,\vrt,Q)
  \right)^\ast\,
  \sigmadip^{(q)}(\rt,W)\,
  \psi_{\lambda\lambda'}^{(q)\,\pm}(\alpha,\vrt,Q)\,,
\\
\label{xslnoq2}
\sigma_L(W,Q^2) &= \sum_q \int \ud^2 r \int_0^1 \ud\alpha \sum_{\lambda,\lambda'}
  \left(\psi_{\lambda\lambda'}^{(q)\,L}(\alpha,\vrt,Q)
  \right)^\ast\,
  \sigmadip^{(q)}(\rt,W)\,
  \psi_{\lambda\lambda'}^{(q)\,L}(\alpha,\vrt,Q)\,.
\end{align}
Here $\alpha$ is the longitudinal momentum fraction of the 
$\gamma^\ast$
carried by the quark, $\vrt$ is the vector in transverse position space
from the antiquark to the quark, $r=|\vrt|$, 
and $\lambda$ and $\lambda'$ are the 
helicities of $q$ and $\bar q$, respectively. 
The total cross section for the scattering of the $q\bar q$ pair on the proton
is denoted by $\sigmadip^{(q)}$,
the wave functions for transversely and longitudinally polarised
virtual photons by $\psi_{\lambda\lambda'}^{(q)\,\pm}$ and
$\psi_{\lambda\lambda'}^{(q)\,L}$, respectively.
For the explicit form of these wave functions see appendix \ref{appA}. 
A sum over all contributing quark flavours $q$ is to be performed.

However, the standard representation of the photon wave functions in 
terms of longitudinal momentum fraction $\alpha$ and transverse 
position $\vrt$ is not suitable for a discussion of Ioffe-time distributions. 
To study them we have to go to longitudinal {\em and} transverse 
momentum space. There we can directly read off the energy mismatch 
between the $\gamma^*$ and the $q\bar{q}$ pair, both for the 
initial and the final states. Let us, therefore, consider first the 
splitting of the initial $\gamma^*$ into a $q\bar{q}$ pair in 
momentum space. We have 3-momentum but not energy conservation 
at this splitting. Taking this into account we define 
the $4$-momentum of the quark, $k$, and that of the antiquark, $k'$, by 
\begin{align}\label{kquarkdecomp}
 \nvec{k} &= \alpha \nvec{q} + \vkt\,,&
 k^0 &= \sqrt{\nvec{k}^2+m_q^2}\,,\notag\\
 \nvec{k}' &= (1-\alpha) \nvec{q} - \vkt\,,&
 k'^0 &= \sqrt{\nvec{k}'^2+m_q^2}\,.
 \end{align}
Here, in essence, $\alpha \in [0,1]$. The precise $\alpha$-range is 
discussed in appendix \ref{appB}. 
Then the photon wave functions in momentum space are easily 
calculated. For a derivation see for example \cite{Ewerz:2006vd}. 
We give the results in appendix \ref{appA} 
both in the -- in leading order in $\alpha_{\rm s}$ and $\alpha_{\rm em}$ -- 
exact and in the approximate form which is usually used in 
the dipole model fits. 
The approximation involves in particular neglecting terms such as
$(\vkt^2 +m_q^2)/ (\alpha^2 \nvec{q}^2)$ and 
$(\vkt^2+m_q^2) / ((1-\alpha)^2 \nvec{q}^2)$ with respect to 1.
However, for some given $\abs{\nvec{q}}$ those terms become non-negligible
if $\kt=|\vkt |$ is large or if $\alpha$ is close to $0$ or $1$. 
If relevant contributions to some observable depend on the photon wave functions
in this kinematical region the above approximation could become invalid.
This is potentially important when considering distributions in the Ioffe times,
in particular for short Ioffe times.
We shall calculate the distributions with and without the above 
approximation in order to quantify this effect.

The photon wave functions in transverse position space are related by
a Fourier transformation to their momentum space representations 
\begin{equation}\label{psifourier}
\psi_{\lambda\lambda'}^{(q)\,\pm,L}(\alpha,\vrt,Q) =
  \int \frac{\ud^2 \kt }{(2\pi)^2}  \, e^{i \vkt \vrt}
  \tilde{\psi}_{\lambda\lambda'}^{(q)\,\pm,L}(\alpha,\vkt,Q)\,.
\end{equation}
The next step is to insert the representation \eqref{psifourier} for 
the wave functions for both the initial and final state photon into 
\eqref{xstnoq2} and \eqref{xslnoq2}. This gives 
\begin{align}
\label{xstmomspace}
\sigma_T(W,Q^2) &= \sum_{q,\lambda,\lambda'}
  \int\! \ud\alpha
  \frac{\ud^2\kbt}{(2\pi)^2}
  \frac{\ud^2\kat}{(2\pi)^2}
  \left(\tilde{\psi}^{(q)\,\pm}_{\lambda\lambda'}(\alpha,\vkbt,Q)
  \right)^\ast
  \tilde{\sigmadip}^{(q)}(\vkat-\vkbt, W)
  \tilde{\psi}^{(q)\,\pm}_{\lambda\lambda'}(\alpha,\vkat,Q)\,,
\\
\label{xslmomspace}
\sigma_L(W,Q^2) &= \sum_{q,\lambda,\lambda'}
  \int\! \ud\alpha
  \frac{\ud^2 \kbt }{(2\pi)^2}
  \frac{\ud^2 \kat }{(2\pi)^2}
  \left(\tilde{\psi}^{(q)\,L}_{\lambda\lambda'}(\alpha,\vkbt,Q)
  \right)^\ast
 \tilde{\sigmadip}^{(q)}(\vkat-\vkbt, W)
  \tilde{\psi}^{(q)\,L}_{\lambda\lambda'}(\alpha,\vkat,Q)\,.
\end{align}
Here we denote the Fourier transform of the dipole-proton cross section by 
\begin{equation}
\label{xsdipfourier}
\tilde{\sigmadip}^{(q)}(\vkt, W) =
  \int\ud^2\rt\, e^{i \vkt \vrt}\,\sigmadip^{(q)}(\rt,W)\,.
\end{equation}

The four-momenta of the quark and antiquark in the initial state 
dipole are given in \eqref{kquarkdecomp}. For the quark and 
antiquark in the final state dipole we denote the 4-momenta by 
$\bar{k}$ and $\bar{k}'$, respectively. They are obtained from 
\eqref{kquarkdecomp} with the replacements $k\to \bar{k}$, 
$k' \to \bar{k}'$, and $\vkt \to \vktbar$. Note that 
$\alpha$ stays the same. 

We recall now the definition of the 
energy mismatches $\dEa$ for the initial $\gamma^\ast$ splitting to
$q\bar q$ and $\dEb$ for the final $q\bar q$ fusing to $\gamma^\ast$: 
\begin{alignat}{2}
\label{dEadef}
\dEa
  &= \ka^0 + \ka'^0 - q^0\,,
\\
\label{dEbdef}
\dEb
  &= \kb^0 + \kb'^0 - q^0\,.
\end{alignat}
The corresponding Ioffe times are
\begin{align}
\label{taudef}
\taua &= \frac{1}{\dEa}\,, \\
\label{taubdef}
\taub &= \frac{1}{\dEb}\,.
\end{align}
We have $\Delta E \ge 0$ and $\Delta \bar{E} \ge 0$ which implies 
also $\tau \ge 0$ and $\bar{\tau} \ge 0$. 

From \eqref{xstmomspace} and \eqref{xslmomspace} we see that the 
cross sections $\sigma_T$ and $\sigma_L$ involve the superpositions
of amplitudes of various $\dEa$ in the initial and various $\dEb$ in the final
state. 
Therefore, we define the joint distributions in $\etaa=\dEa$ and $\etab=\dEb$ by 
\begin{equation}
\label{xsdetadetab}
\begin{split}
\frac{\partial^2 \sigma_{T,L} (W,Q^2,\eta,\bar{\eta})}{\partial\etaa\,\partial\etab} =
  \sum_q \sum_{\lambda,\lambda'} \int\! \ud\alpha &
  \int\! \frac{\ud^2 \kbt}{(2\pi)^2}
  \int\!  \frac{\ud^2 \kat}{(2\pi)^2}
  \left( \tilde{\psi}^{(q)\,\pm,L}_{\lambda\lambda'}(\alpha,\vkbt,Q)
  \right)^\ast\,
  \delta(\etab-\dEb)
\\ 
&\times \tilde{\sigmadip}^{(q)}(\vkat-\vkbt,W)\,
  \delta(\etaa-\dEa)\,
  \tilde{\psi}^{(q)\,\pm,L}_{\lambda\lambda'}(\alpha,\vkat,Q)\,.
\end{split}
\end{equation}
Note that these distributions are real-valued (see \eqref{A3e}) 
but for $\etaa\neq \etab$ they need not be positive. 
For specific dipole models we can now calculate $\tilde{\sigmadip}^{(q)}$ 
and evaluate the two-dimensional distributions \eqref{xsdetadetab}. 
However, one-dimensional distributions are more easily visualised. 
Thus, in the following we shall study how the sum
of $\dEa$ and $\dEb$
\begin{equation}\label{dEpdef}
\dEp = \dEa + \dEb 
\end{equation}
is distributed. We, therefore, define 
\begin{align}
\label{xstldEp}
\frac{\partial\sigma_{T,L}(W,Q^2,\dEp)}{\partial\dEp}
&= \int_0^\infty\! \ud\eta \int_0^\infty\! \ud \bar{\eta}\,
  \frac{\partial^2 \sigma_{T,L}(W,Q^2,\eta,\bar{\eta})}{\partial\eta\,\partial\bar{\eta}}\,
  \delta(\dEp-\eta-\bar{\eta}) 
\end{align}
and the Ioffe time corresponding to $\dEp$ by 
\begin{equation}\label{taupdef}
\tau_{+} = \frac{1}{\dEp} =\frac{\tau \bar{\tau}}{\tau + \bar{\tau}}
\,.
\end{equation}
The quantity $2 \tau_+$ is called the `harmonic mean' of $\tau$ 
and $\bar{\tau}$ \cite{encyclopedia}. 
From \eqref{taupdef} we get then $\tau \ge \tau_+$ and 
$\bar{\tau} \ge \tau_+$. That is, for given $\tau_+$ only individual 
Ioffe times $\tau$ and $\bar{\tau}$ which are {\em larger} or equal to 
$\tau_+$ contribute. At this point we may also note that Ioffe in his original 
paper \cite{Ioffe:1969kf} considered the time between the initial $\gamma^*$ 
fluctuating into `hadronic stuff' and this `stuff' fluctuating back to the final 
$\gamma^*$. In our calculation this corresponds to the time $\tau + \bar{\tau}$. 
For given $\tau_+$ we have $\tau + \bar{\tau} \ge 4 \tau_+$. 

In the next section we shall present numerical results for normalised 
distributions in $\ln \Delta E_+$ respectively $\ln \tau_+= - \ln \Delta E_+$. 
First we shall study these distributions for $F_2$, 
\begin{equation}
\label{distE+F2}
\frac{\Delta E_+}{F_2} \frac{\partial F_2}{\partial \Delta E_+} = 
\frac{\Delta E_+}{\sigma_T + \sigma_L} 
\left( \frac{\partial \sigma_T}{\partial \Delta E_+} 
+ \frac{\partial \sigma_L}{\partial \Delta E_+} \right)
\,.
\end{equation}
It is also of interest to study the distributions in the Ioffe time $\tau_+$ 
and in $\Delta E_+$ for the contributions of individual quark flavours 
to $\sigma_T$ and $\sigma_L$. 
For this we define $\sigma_{T}^{(q)}$, $\sigma_{L}^{(q)}$, 
and  $\partial^2 \sigma_{T,L}^{(q)}/\partial \eta \partial \bar{\eta}$ 
as in \eqref{xstmomspace}, \eqref{xslmomspace}, and \eqref{xsdetadetab}, 
respectively, but omitting the sum over the quark flavours $q$ on the r.h.s. 
Then we define $\partial \sigma_{T,L}^{(q)}/\partial \Delta E_+$ 
in analogy to \eqref{xstldEp}. 
The corresponding normalised distributions in $\ln \Delta E_+$ 
are given by 
\begin{equation}
\label{dEpsigTL}
\frac{\Delta E_+}{\sigma_{T,L}^{(q)}}
\frac{\partial \sigma_{T,L}^{(q)}}{\partial \Delta E_+} \,.
\end{equation}

\section{Results}
\label{sec:results}

In this section we present results for the distributions in $\Delta E_+$ 
and in Ioffe times $\tau_+$ in DIS. To evaluate \eqref{distE+F2} 
and \eqref{dEpsigTL} we have to choose a specific dipole-model 
fit to the data since we need the values for the dipole-proton cross 
sections $\hat{\sigma}^{(q)}$; see \eqref{xstnoq2}, \eqref{xslnoq2}, 
and \eqref{xsdipfourier}. 
In the following we choose to work with the model constructed by 
Golec-Biernat and W\"usthoff \cite{GolecBiernat:1998js}. 
This GBW model describes the $F_2$ data from HERA quite well.
Whether it also describes the longitudinal structure function $F_L$ is 
not yet clear since the first $F_L$ measurements from HERA \cite{:2008tx} are 
not precise enough to draw firm conlusions.
The dipole cross section of this model is given for quark flavour $q$ by
\begin{equation}\label{xsgbw}
\hat{\sigma}_{\mathrm{GBW}}^{(q)}(r,x)
  = \sigma_0 \left[1-\exp\left(-\frac{r^2}{4 r_0^2(\tilde{x})}\right)\right],
\end{equation}
where 
\begin{equation}
\label{GBWparam}
r_0^2(\tilde{x}) = \frac{1}{Q_0^2}\left( \frac{\tilde{x}}{x_0} \right)^\lambda\,,
\qquad
\tilde{x} = x \left(1 + \frac{4 m_q^2}{Q^2} \right)\,.
\end{equation}
We choose their parameter set which includes the charm quark, 
$Q_0=1~\mbox{GeV}$, $\sigma_0 = 29.12 ~\mbox{mb}$, $\lambda = 0.277$, 
$x_0= 0.41\cdot 10^{-4}$, $m_q=0.14~\mbox{GeV}$ for $q=u,d,s$, 
and $m_c=1.5~\mbox{GeV}$. 
The $b$-quark contributions are neglected in this fit. 

The GBW dipole cross section $\hat{\sigma}_{\rm GBW}^{(q)}$ 
depends on $r$ and Bjorken-$x$, and therefore not only on $r$ and 
$W$ but also on $Q^2$, see \eqref{defkin}. 
As was discussed in \cite{Ewerz:2006vd,Ewerz:2006an} the natural and 
-- in our opinion -- correct 
energy variable for the dipole cross section $\hat{\sigma}^{(q)}$ is $W$, 
and $\hat{\sigma}^{(q)}$ should actually be independent of $Q^2$. 
A dependence on $x$ requires additional assumptions which appear 
difficult to justify. 
We will discuss the problem of choosing the correct energy variable 
in more detail elsewhere. For the present considerations, however, 
this issue how the $\hat{\sigma}^{(q)}$ depend on the kinematic 
variables is not of prime relevance since we only need a dipole model 
fit to the data at given values of these variables. 

Now it is relatively straightforward to calculate the Fourier transform 
of $\hat{\sigma}_{\mathrm{GBW}}^{(q)}$, to insert it as well as the 
expressions for the $\gamma^*$ wave functions from appendix \ref{appA} 
in \eqref{xstmomspace}, \eqref{xslmomspace} and \eqref{xsdetadetab}, 
and to obtain the numerical 
results for the distributions in $\ln \Delta E_+=-\ln \tau_+$. 
The calculational details are presented 
in appendix \ref{appB}. The results for the normalised $F_2$ distribution 
\eqref{distE+F2} are shown for $\gamma^* p $ 
c.\,m.\ energies $W=70$, $150$ and $220~\mbox{GeV}$ and a number 
of $Q^2$ values in figure \ref{fig:ndF2}. 
\FIGURE[ht]{
\includegraphics[width=\textwidth]{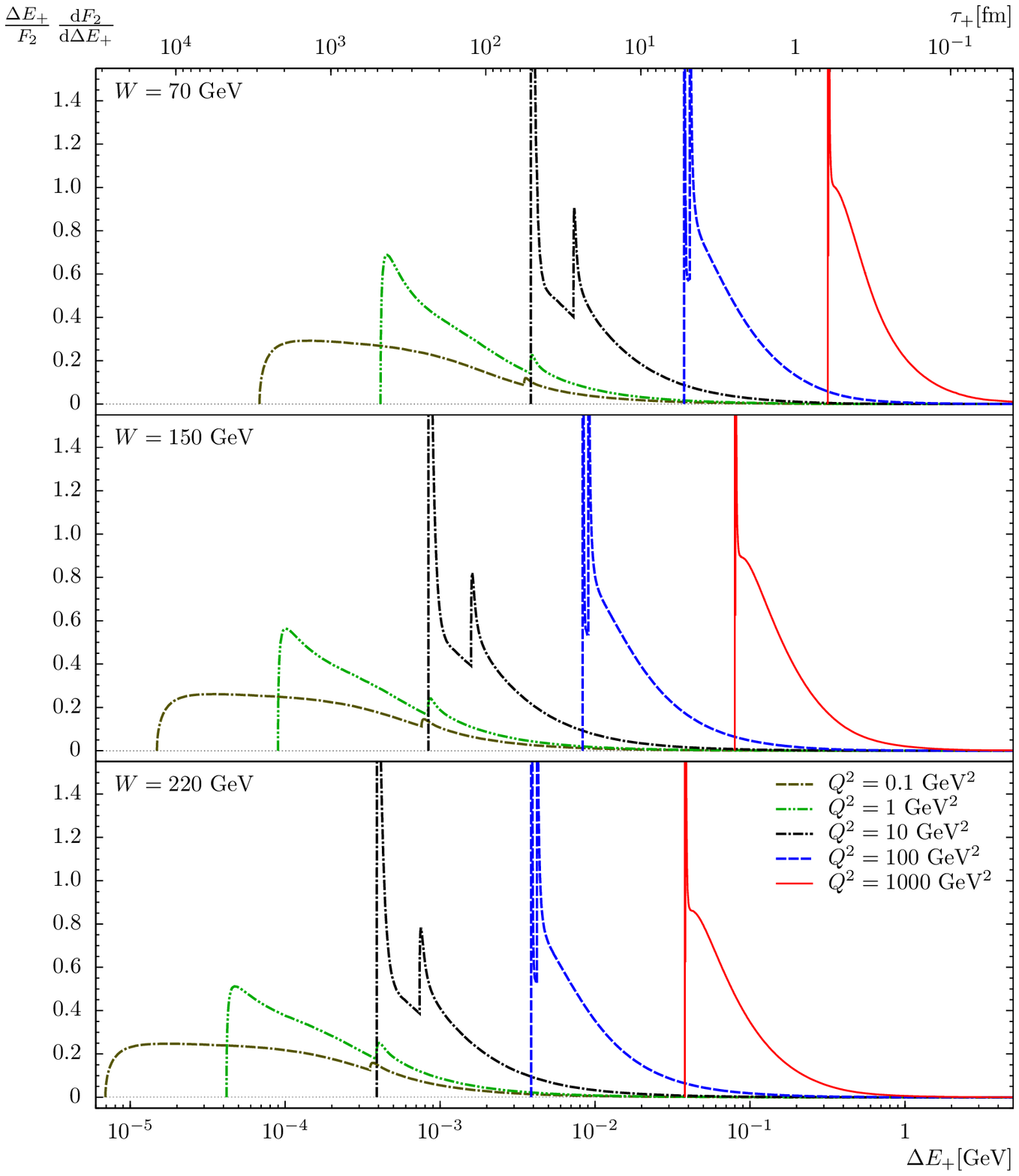}
\caption{
The normalised $F_2$ distribution \eqref{distE+F2} in the joint 
dipole-energy mismatch $\dEp$ for the GBW model with charm.
Supplementary to the $\dEp$ values the corresponding values for the
Ioffe times $\tau_+$ are denoted for the abscissa.
\label{fig:ndF2}
}
}
In figure \ref{fig:ndxq} we present the normalised 
distributions in $\ln \Delta E_+= - \ln \tau_+$ for $\sigma_T^{(q)}$ 
and $\sigma_L^{(q)}$ for the individual quark flavours separately, 
here for the energy $W=150~\mbox{GeV}$.
\FIGURE[ht]{
\includegraphics[width=\textwidth]{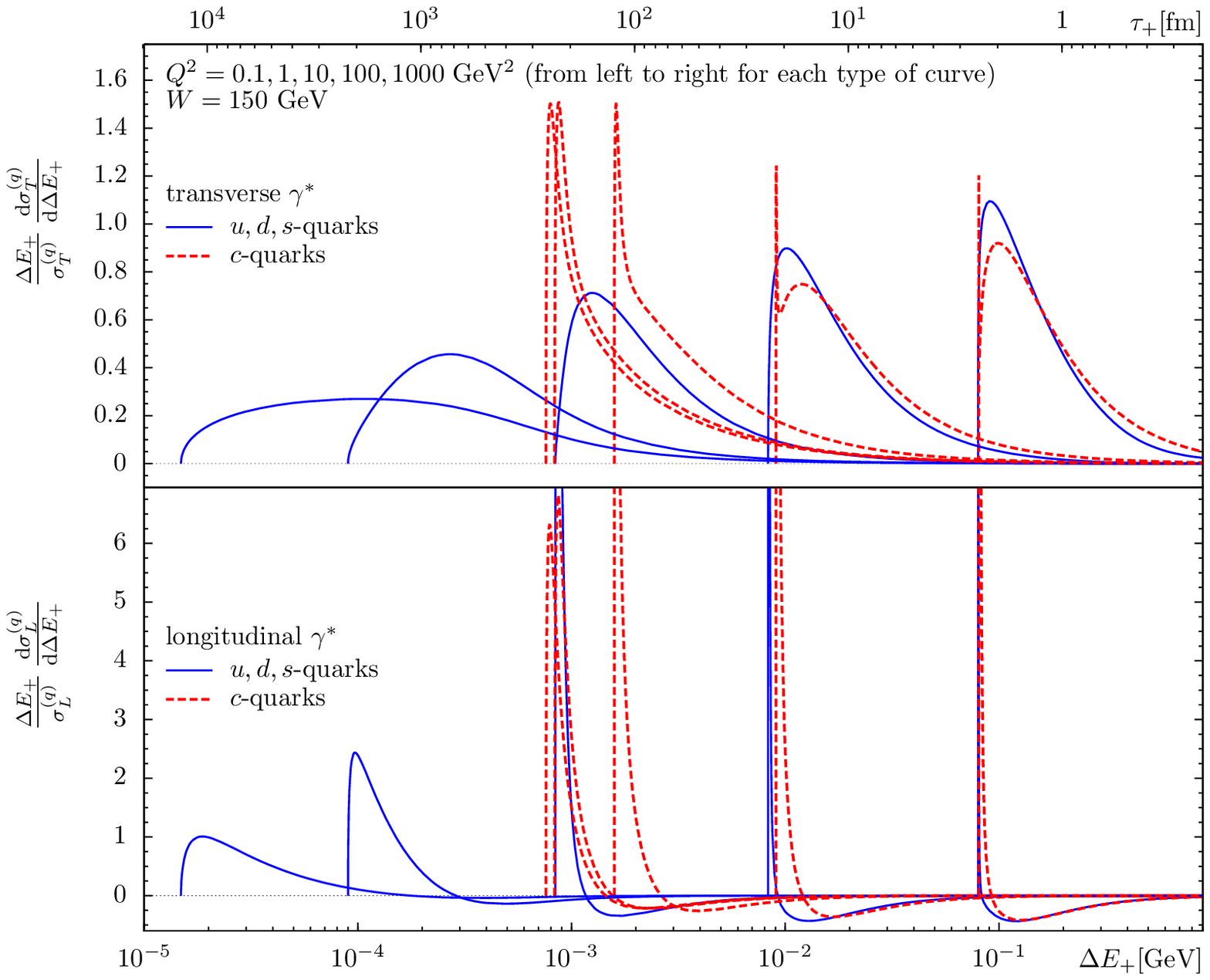}
\caption{
The normalised $\sigma_T^{(q)}$ and $\sigma_L^{(q)}$ distributions
\eqref{dEpsigTL} in the joint dipole-energy mismatch $\dEp$ respectively Ioffe time
$\tau_+$ for the GBW model with charm.
\label{fig:ndxq}
}
}

In the calculations for figures \ref{fig:ndF2} and \ref{fig:ndxq} 
the full expressions \eqref{psitkgen} and \eqref{psilkgen} are used
for the photon wave functions rather than their high energy and 
small-$k_T$ approximation.
Also, we integrate over the full range in $\alpha$ which is slightly bigger
than $[0,1]$, see \eqref{alphaminmax}.
We have studied the effect of the high energy approximation 
\eqref{psitk}-\eqref{psil} for the photon wave functions as well as of restricting the
$\alpha$ integration range to $[0,1]$.
The Ioffe-time distributions in figures \ref{fig:ndF2} and \ref{fig:ndxq} 
are essentially
unaltered when performing the $\alpha$ integration only over $[0,1]$.
The $\alpha$ range specified by \eqref{alphaminmax} is only slightly larger
than $[0,1]$ and no sufficiently strong enhancement of the integrand
compensates for this fact in the kinematical ranges considered here.
As expected, effects in the Ioffe-time distributions from the high energy 
and small-$k_T$ approximation for the photon wave functions 
become larger for increasing $\dEp$.
These deviations would alter the curves of figures \ref{fig:ndF2} and
\ref{fig:ndxq} to an amount that is clearly visible only in the
high $Q^2$ and small $W$ cases.
However, for all distributions they are still small enough to be safely
omitted from our further discussions here.
We stress that this result is not obvious, since the interplay of
the high energy and small-$k_T$ 
approximation and the dipole energy mismatch is non-trivial
in the longitudinal-momentum endpoint regions. 

Let us now discuss the main features of the curves in figures 
\ref{fig:ndF2} and \ref{fig:ndxq}. We first note that due to the 
normalisation of the curves in figure \ref{fig:ndxq}  
one cannot deduce from them the relative importance of light quarks 
and the $c$ quark in $F_2$. This information is obtained from 
figure \ref{fig:ndF2} where the secondary peaks in the individual 
curves are due to the $c$-quark contributions.
The peaks seen in figures \ref{fig:ndF2} and \ref{fig:ndxq} 
are not infinitely sharp.
We analysed their structure using an enlarged $\dEp$ scale and find
them to be smooth finite maxima in all cases.

The curves in figure \ref{fig:ndxq} clearly exhibit the respective thresholds
due to the kinematical lower bounds for the various quark flavours $q$,
\begin{equation}
\label{8cc}
\Delta E_{+,\text{min}} 
= 2 \left( \sqrt{ \nvec{q}^2 + 4 m_q^2} - q^0 \right) 
= 2 \left(  \sqrt{ \nvec{q}^2 + 4 m_q^2} - \sqrt{  \nvec{q}^2 -Q^2} \right) \,,
\end{equation}
see \eqref{dEmintot} and \eqref{dEpdef}.
The quantities $\Delta E_{+,\text{min}}$ are shown in the left graph of 
figure \ref{fig:deltaemin} for light quarks as functions of $Q^2$ for 
$W=70$, $150$ and $220~\mbox{GeV}$.
\FIGURE[ht]{
\includegraphics[width=\textwidth]{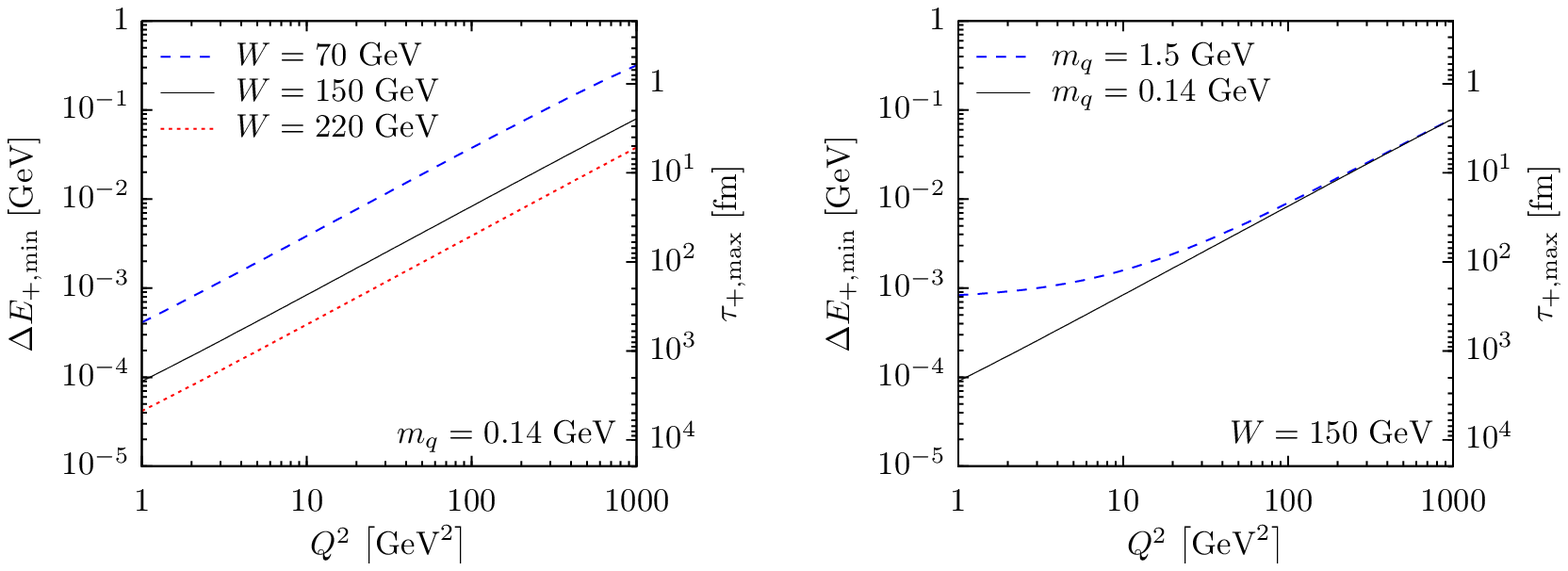}
\caption{Minimal energy mismatch $\dE_{+,\text{min}}$ and corresponding
maximal Ioffe time $\tau_{+,\text{max}}$ of the 
$\gamma^*$--dipole transition as a function of the 
photon virtuality $Q^2$.
The curves are for different values of the energy $W$ in the case of light
quarks (left graph) and for different quark masses in the case
$W=150~\mbox{GeV}$ (right graph). \label{fig:deltaemin}
}
}
In the right graph of figure \ref{fig:deltaemin} 
we show  $\Delta E_{+,\text{min}}$ as a function of $Q^2$ for 
$W=150\,\mbox{GeV}$ for the light quark mass and for the charm mass. 
From figure \ref{fig:ndxq} we see that 
the $\dEp$ distributions for transverse $\gamma^\ast$ polarisation are
significantly broader than those for longitudinal $\gamma^\ast$ polarisation.
For both polarisation types the distributions become narrower for increasing
values of $Q^2$.
The distributions for longitudinal polarisation have negative parts which 
is alright, see the discussion after~\eqref{xsdetadetab}. 

The normalised distributions in $\tau_+$ respectively $\dEp$ shown in 
figure \ref{fig:ndF2} are obtained from the DIS small-$x$ data from HERA, 
using a specific dipole-model fit. In order to test the model dependence we have 
also investigated another fit of the GBW type and find compatible results. 

From figure \ref{fig:ndF2} we can clearly see that the Ioffe times decrease 
with increasing $Q^2$ at fixed $W$. At fixed $Q^2$ the Ioffe times increase 
with increasing $W$. The distributions are always rather broad. At 
$W=70~\mbox{GeV}$ and $Q^2=100~\mbox{GeV}^2$, for example, 
the upper cutoff for the Ioffe time $\tau_+$ is 5 fm and, indeed, the 
peaks of the distribution are there. But we have also significant contributions 
for $\tau_+$ down to 1 fm. 
As stressed already in the introduction, for the dipole picture to be 
a valid description of DIS the Ioffe times have to be appreciably larger than 
the typical interaction time $\tau_{\rm had}$ of the dipole with the proton. 
As an estimate of the latter we take the electromagnetic radius of the 
proton $\tau_{\rm had} \approx r_p \approx 1~\mbox{fm}$. 
A necessary condition for both dipole lifetimes $\taua$ and $\taub$ 
(see \eqref{taudef}, \eqref{taubdef}) 
to be much larger than $\tau_{\rm had}$ is thus $\dEp \ll 1/\tau_{\rm had}$ 
or, equivalently,
\begin{align}
\label{sepcondioffe}
\tau_{+} &\gg \tau_{\rm had} \approx 1~\text{fm}\,.
\end{align}
If this condition is violated for a relevant portion of the $\tau_+$ 
distribution of $F_2$ the applicability of the dipole model is questionable 
for the kinematical point under consideration. 

From figure \ref{fig:ndF2} we see that for $W=70~\mbox{GeV}$ 
the typical Ioffe times $\tau_{+}$ are reasonably above $1$~fm for all 
$Q^2\leq 100$~GeV$^2$. There, the 
necessary separation condition \eqref{sepcondioffe} is satisfied.
However, for $Q^2=1000$~GeV$^2$ we see that 
the separation condition \eqref{sepcondioffe} is clearly violated, 
since the complete $\tau_+$ distribution lies at $\tau_+ < 1~\mbox{fm}$. 
For $W=150~\mbox{GeV}$ the distributions are shifted to larger 
values of $\tau_+$. But still for $Q^2=1000~\mbox{GeV}^2$ the 
$\tau_+$ distribution only starts at $\tau_+ \approx 2.5~\mbox{fm}$ 
and extends well below $\tau_+=1~\mbox{fm}$. 
For $W=220~\mbox{GeV}$ and $Q^2=1000~\mbox{GeV}^2$ the 
situation is similar with the maximal $\tau_+\approx 5~\mbox{fm}$. 
Therefore, we must conclude from our study of Ioffe-time distributions 
in DIS that the basic separation condition of time scales \eqref{sepcondioffe} 
is violated in the HERA energy range $W=70$ to $220~\mbox{GeV}$, 
for $Q^2$ values of hundred to several hundred GeV$^2$. 

\section{Conclusions}
\label{sec:conclusions}

We have calculated Ioffe-time distributions for DIS in the HERA energy 
range in the framework of the dipole model. As a convenient fit to the 
data we used the Golec-Biernat-W\"usthoff parametrisation of the 
dipole-proton cross section. We have obtained Ioffe-time distributions 
for $\gamma^* p$ c.\,m.\ energies $W=70$, $150$ and $220~\mbox{GeV}$ 
and $Q^2$ values ranging from $0.1$ to $1000~\mbox{GeV}^2$. 
A basic requirement for the dipole model to make sense is that the Ioffe 
times at the kinematic point considered are much larger than the hadronic 
timescale $\tau_{\rm had} \approx 1~\mbox{fm}$. 
We find that typical Ioffe times are large with respect
to this hadronic scale for $Q^2\leq 100$~GeV$^2$,
such that no inconsistencies arise there. 
However, at photon virtuality $Q^2=1000$~GeV$^2$ 
typical Ioffe times are of similar order as or smaller than 
the hadronic scale, which violates the standard assumption for the 
validity of the dipole picture. 
Thus, we find that in the HERA energy range the dipole picture starts 
to lack physical justification for $Q^2$ values in the hundred to 
several hundred GeV$^2$ range. 

Note that we come to this conclusion here using a GBW fit to the 
HERA data where the dipole-proton cross section is assumed to depend 
on $r$ and Bjorken $x$. In \cite{Ewerz:2007md} we have investigated 
the limits of applicability of the dipole model to the HERA data using 
the -- in our opinion correct -- dependence of the dipole-proton cross 
section on $r$ and $W$. From figure 9 of \cite{Ewerz:2007md} 
we see that, nevertheless, we found a very similar $Q^2$ range for the 
applicability of the dipole model. The upper limits for $Q^2$ as 
obtained from this figure range from about $120~\mbox{GeV}^2$ 
at $W=70~\mbox{GeV}$ to about $180~\mbox{GeV}^2$ at 
$W=220~\mbox{GeV}$. These are strict bounds in the sense that the 
data cannot be fitted for a larger $Q^2$ range with a non-negative 
dipole-proton cross section depending on $r$ and $W$. 
In our present work we find that even if a fit to the data with a 
dipole-proton cross section depending on $r$ and Bjorken $x$ is 
possible also for higher values of $Q^2$ its physical meaning may 
become questionable there since the Ioffe times become too short. 
For very small $Q^2$ values, $Q^2 \le 2~\mbox{GeV}^2$ say, 
the Ioffe times shown in figure \ref{fig:ndF2} are very large. 
Nevertheless, this does not immediately imply that the 
standard dipole model is without problems there. As discussed in 
\cite{Ewerz:2006vd,Ewerz:2006an,Ewerz:2007md} the lowest 
order expressions for the photon wave functions are expected to 
become unreliable in this kinematic region. 

In summary, we have calculated `experimental' Ioffe-time distributions 
from the \mbox{small-$x$} HERA data as described by a GBW dipole-model 
fit. We have studied their dependence on the energy $W$ and 
on the photon virtuality $Q^2$. The Ioffe-time distributions of the
cross sections are found to be rather broad for transversely and very narrow
for longitudinally polarised virtual photons.
Accordingly, the Ioffe-time distributions of $F_2$ are always rather broad.

\section*{Acknowledgements}
We would like to thank M.\ Diehl for useful discussions. 
C.\,E.\ was supported by the Alliance Program of the
Helmholtz Association (HA216/EMMI) and by the 
Deutsche Forschungs\-gemeinschaft, project Sh 92/2-1. 
A.\,v.\,M.\ was supported by the Schweizer Nationalfonds.
The support of this work by the Deutsche 
For\-schungsgemeinschaft under project number Na 296/4-1 
is gratefully acknowledged. 

\begin{appendix}

\boldmath
\section{Photon wave functions and energy mismatch $\Delta E$}
\unboldmath
\label{appA}

In this appendix we collect the formulae of the dipole model which are 
relevant for our calculations. The assumptions needed to arrive at the standard 
dipole picture of DIS are spelled out in detail in section 6.2 of \cite{Ewerz:2006vd}. 
One obtains the following -- in leading order in $\alpha_{\rm s}$ and
$\alpha_{\rm em}$ exact -- 
expressions for the momentum-space 
wave functions of the virtual photon from (52) and (58) of \cite{Ewerz:2006vd}, 
\begin{align}
\label{psitkgen}
\tilde{\psi}^{(q)\,\pm}_{\lambda\lambda'}(\alpha,\vkat,Q) =&
  \mp \frac{N}{\sqrt{2}} Q_q \frac{1}{\dE}\frac{\abs{\nvec{q}}}{2\pi k^0 2 k'^0}
  \frac{2}{\sqrt{k^0+m_q}\sqrt{k'^0+m_q}}
\notag\\ 
&\times
  \Big[
     \pm \left( (k^0+m_q)(k'^0+m_q)-\alpha(1-\alpha)\nvec{q}^2 \right)
          \delta_{\lambda,\lambda'}\delta_{\lambda,\pm\frac{1}{2}}
\\ 
&\quad\quad
    +e^{\pm i \phi_k} \kt \abs{\nvec{q}} \delta_{\lambda,-\lambda'}
    \big( \alpha\delta_{\lambda,\pm\frac{1}{2}}
           - (1-\alpha)\delta_{\lambda,\mp\frac{1}{2}} \big)
    \pm e^{\pm i 2 \phi_k} \kt^2
      \delta_{\lambda,\lambda'} \delta_{\lambda,\mp\frac{1}{2}}
  \Big]\,,
\notag
\\
\label{psilkgen}
\tilde{\psi}^{(q)\,L}_{\lambda\lambda'}(\alpha,\vkat,Q) =&
  -N Q_q \frac{1}{\dE}\frac{\abs{\nvec{q}}}{2\pi k^0 2 k'^0}
  \frac{\abs{\nvec{q}}-q^0}{Q}
  \frac{1}{\sqrt{k^0+m_q}\sqrt{k'^0+m_q}}
\notag\\ &\times
  \Big[ \big( \kt^2 + (k^0+m_q+\alpha\abs{\nvec{q}})
                       (k'^0+m_q+(1-\alpha)\abs{\nvec{q}}) \big)
           \delta_{\lambda,-\lambda'}
\notag\\ &\quad\quad
         + e^{-i (\sign\lambda )\phi_k}(\sign\lambda)
           (k^0-k'^0-(1-2\alpha)\abs{\nvec{q}})\kt \delta_{\lambda,\lambda'}
  \Big]
\,.
\end{align}
Here $q$ denotes the quark flavours, $Q_q$ their charges in units of the proton 
charge $e=\sqrt{4\pi\alpha_{\rm em}}$. The number of colours is $N_c=3$.
The momenta $k$ and $k'$ of the quark and antiquark, respectively, 
are defined in \eqref{kquarkdecomp}, and we have $k_T = |\vkat|$ and 
$\phi_k = \arg (k_{T1} + i k_{T2})$. The normalisation factor is 
\begin{equation}
N = -2\sqrt{N_c \pi}\,e \,\sqrt{\alpha(1-\alpha)} \,.
\end{equation}

Note that the wave functions satisfy the following relations 
\begin{align}
\label{A3a}
\left( 
\tilde{\psi}^{(q)\,\pm}_{\lambda\lambda'}(\alpha,\vkat,Q) 
\right)^* 
&= 
\tilde{\psi}^{(q)\,\pm}_{\lambda\lambda'}(\alpha,\vkat^R,Q) \,,
\\
\label{A3b}
\left(
\tilde{\psi}^{(q)\,L}_{\lambda\lambda'}(\alpha,\vkat,Q) 
\right)^*
&=
\tilde{\psi}^{(q)\,L}_{\lambda\lambda'}(\alpha,\vkat^R,Q) \,,
\end{align}
where 
\begin{equation}
\label{A3c}
\vkat^R= 
\left(\begin{array}{c}k_{T1} \\ -k_{T2}\end{array}\right) \,.
\end{equation}

With \eqref{A3a} and \eqref{A3b} we can show that the distributions 
defined in \eqref{xsdetadetab} are real-valued. Indeed, from 
\eqref{xsdipfourier} we find that $\tilde{\sigmadip}^{(q)}(\vkt, W)$ 
is independent of the direction of $\vkt$ and, therefore, 
\begin{align}
\label{A3d}
\left( \tilde{\sigmadip}^{(q)}(\vkt, W) \right)^* 
&= \tilde{\sigmadip}^{(q)}(-\vkt, W)
\notag
\\
&= \tilde{\sigmadip}^{(q)}(\vkt^R, W) \,.
\end{align}
We then get from \eqref{xsdetadetab} 
\begin{align}
\label{A3e}
\left(
\frac{\partial^2 \sigma_{T,L} (W,Q^2,\eta,\bar{\eta})}{\partial\etaa\,\partial\etab} 
\right)^*
&= \sum_q \sum_{\lambda, \lambda'} \int\! \ud\alpha 
\int\! \frac{\ud^2 \kbt}{(2\pi)^2} \int\!  \frac{\ud^2 \kat}{(2\pi)^2} \,
\tilde{\psi}^{(q)\,\pm,L}_{\lambda\lambda'}(\alpha,\vkbt,Q) \,
\delta(\etab-\dEb) 
\notag \\
& \hspace*{1cm}
\times \left( \tilde{\sigmadip}^{(q)}(\vkat-\vkbt,W)\right)^* 
\delta(\etaa-\dEa) 
\left(\tilde{\psi}^{(q)\,\pm,L}_{\lambda\lambda'}(\alpha,\vkat,Q) \right)^*
\notag \\
&= \sum_q \sum_{\lambda, \lambda'} \int\! \ud\alpha 
\int\! \frac{\ud^2 \kbt}{(2\pi)^2} \int\!  \frac{\ud^2 \kat}{(2\pi)^2}
\left( \tilde{\psi}^{(q)\,\pm,L}_{\lambda\lambda'}(\alpha,\vkbt^R,Q) \right)^* 
\delta(\etab-\dEb) 
\notag \\
& \hspace*{1cm}
\times \tilde{\sigmadip}^{(q)}(\vkat^R-\vkbt^R,W) \,
\delta(\etaa-\dEa)  \,
\tilde{\psi}^{(q)\,\pm,L}_{\lambda\lambda'}(\alpha,\vkat^R,Q) 
\notag \\ 
&= 
\frac{\partial^2 \sigma_{T,L} (W,Q^2,\eta,\bar{\eta})}{\partial\etaa\,\partial\etab} 
\,.
\end{align}

In the high energy and small-$k_T$ limit 
\begin{equation}
\label{highqsmallklim}
\abs{\nvec{q}}\to\infty \,, \quad
Q^2 \ll |\nvec{q}^2| \,,\quad
\frac{\vkat^2+m_q^2}{\alpha^2 \nvec{q}^2} \ll 1\,, \quad
\frac{\vkat^2+m_q^2}{(1-\alpha)^2 \nvec{q}^2} \ll 1\,, 
\end{equation}
the photon wave functions simplify to 
\begin{align}
\tilde{\psi}^{(q)\,\pm}_{\lambda\lambda'}(\alpha,\vkat,Q) &=
  \frac{\sqrt{2 N_c \alpha_{\rm em}} Q_q}{\alpha(1-\alpha)Q^2+\kt^2+m_q^2}
\Big[ m_q \delta_{\lambda,\lambda'}\delta_{\lambda,\pm\frac{1}{2}}
\notag\\
\label{psitk}
&\quad\qquad  \pm\kt e^{\pm i \phi_k} \delta_{\lambda,-\lambda'}
         \left(\alpha\delta_{\lambda,\pm\frac{1}{2}}
              -(1-\alpha)\delta_{\lambda,\mp\frac{1}{2}}\right)
\Big]\,,
\\
\label{psilk}
\tilde{\psi}^{(q)\,L}_{\lambda\lambda'}(\alpha,\vkat,Q) &=
  -\frac{2 \sqrt{N_c \alpha_{\rm em}} Q_q}{\alpha(1-\alpha)Q^2+\kt^2+m_q^2}
\,  \alpha(1-\alpha)Q\,\delta_{\lambda,-\lambda'}
\,.
\end{align}
Inserting the expressions \eqref{psitk}, \eqref{psilk} into 
the Fourier transform \eqref{psifourier} yields the standard formulae 
for the wave functions in longitudinal momentum and transverse 
position space 
(see again \cite{Ewerz:2006vd} for a detailed derivation)
\begin{align}\label{psit}
\psi_{\lambda \lambda'}^{(q)\pm} (\alpha, \vrt,Q)
&=
\frac{\sqrt{N_c}}{\sqrt{2} \pi} \sqrt{\alpha_{\rm em}} \, Q_q
\left\{
\pm \, i e^{\pm i \phi_r }\,\delta_{\lambda', -\lambda} \left[
\alpha \, \delta_{\lambda,\pm \frac{1}{2}} 
- (1 - \alpha) \, \delta_{\lambda,\mp \frac{1}{2}} 
\right] \epsilon_q K_1(\epsilon_q \rt)
\right.
\notag \\
&\quad
\left.
\hspace*{3cm}
+\, m_q \, \delta_{\lambda,\pm \frac{1}{2}} \delta_{\lambda', \lambda}
K_0(\epsilon_q \rt)
\right\}\,,
\\
\label{psil}
\psi_{\lambda \lambda'}^{(q)L} (\alpha, \vrt,Q) &=
- \frac{\sqrt{N_c}}{\pi} \sqrt{\alpha_{\rm em}} \, Q_q
Q \, \alpha (1- \alpha)  \, \delta_{\lambda', -\lambda}  \,
K_0(\epsilon_q \rt) \,.
\end{align}
Here $\phi_r= \arg (r_1 + i r_2)$, 
\begin{equation}
\label{defepsq}
\epsilon_q = \sqrt{\alpha (1-\alpha) Q^2 + m_q^2} \,, 
\end{equation}
and $K_{0,1}$ are the modified Bessel functions. 
Hence we have in the high energy and small-$k_T$ limit 
\eqref{highqsmallklim} for the $\gamma^* p$ 
cross sections the results \eqref{xstnoq2} and \eqref{xslnoq2}, that is, 
\begin{equation}
\label{sigmatdip}
\sigmatot_{T,L}(W,Q^2)=
\sum_q\int \ud^2 \rt 
\int^1_0 \ud \alpha 
\sum_{\lambda,\lambda'}
\left|
\psi^{(q)+,L}_{\lambda\lambda'}(\alpha,\vrt,Q) \right|^2 
\sigmadip^{(q)}(\rt,W) 
\end{equation}
with 
\begin{align}
\label{sumpsitdens}
        \!\sum_{\lambda, \lambda'} \left| 
        \psi_{\lambda \lambda'}^{(q) +} 
          (\alpha, \vrt,Q) \right|^2 
&=
        \frac{N_c}{2 \pi^2} \, \alpha_{\rm em} Q_q^2 
        \left\{  \left[ \alpha^2 + (1-\alpha)^2 \right] 
        \epsilon_q^2 [K_1(\epsilon_q \rt) ]^2 
+ m_q^2 [K_0(\epsilon_q \rt) ]^2 
        \right\},
\\
\label{sumpsildens}
        \!\sum_{\lambda, \lambda'} \left| 
        \psi_{\lambda \lambda'}^{(q) L}(\alpha,\vrt,Q) 
         \right|^2 
&=
        \frac{2 N_c}{\pi^2} \, \alpha_{\rm em} Q_q^2
         Q^2 [\alpha (1-\alpha)]^2 [K_0(\epsilon_q \rt) ]^2  \,.
\end{align}
For the calculation of $\sigma_T$ above we can choose either of the two 
polarisations, $+$ or $-$ in \eqref{sigmatdip}, since they lead to the same 
cross section. We have chosen the $+$ polarisation here. 

Let us now discuss some basic properties of the energy mismatch $\dEa$
\eqref{dEadef} which follow directly from the kinematics of the
$\gamma^* \to q\bar{q}$ splitting.
Similar considerations then apply to $\dEb$ in \eqref{dEbdef}. 
Explicitly we obtain from \eqref{dEadef} and \eqref{kquarkdecomp} 
for quark flavour $q$ 
\begin{align}\label{dEktal}
\dE(k_T,\alpha)
  &= \sqrt{\alpha^2\nvec{q}^2+\kt^2+m_q^2}
   + \sqrt{(1-\alpha)^2\nvec{q}^2+\kt^2+m_q^2}
   - q^0\,.
\end{align}
Figure\ \ref{fig:deltaektal} shows the dependence of $\dE(k_T,\alpha)$ 
on the absolute value $k_T$ of the quark's 
transverse momentum (left graph) and on its longitudinal momentum fraction
$\alpha$ for the case of light quarks (right graph).
We note that $\dE(k_T,\alpha)$ strongly peaks at the longitudinal
momentum endpoints and rises monotonically with the transverse momentum. 
\FIGURE[ht]{
\includegraphics[width=\textwidth]{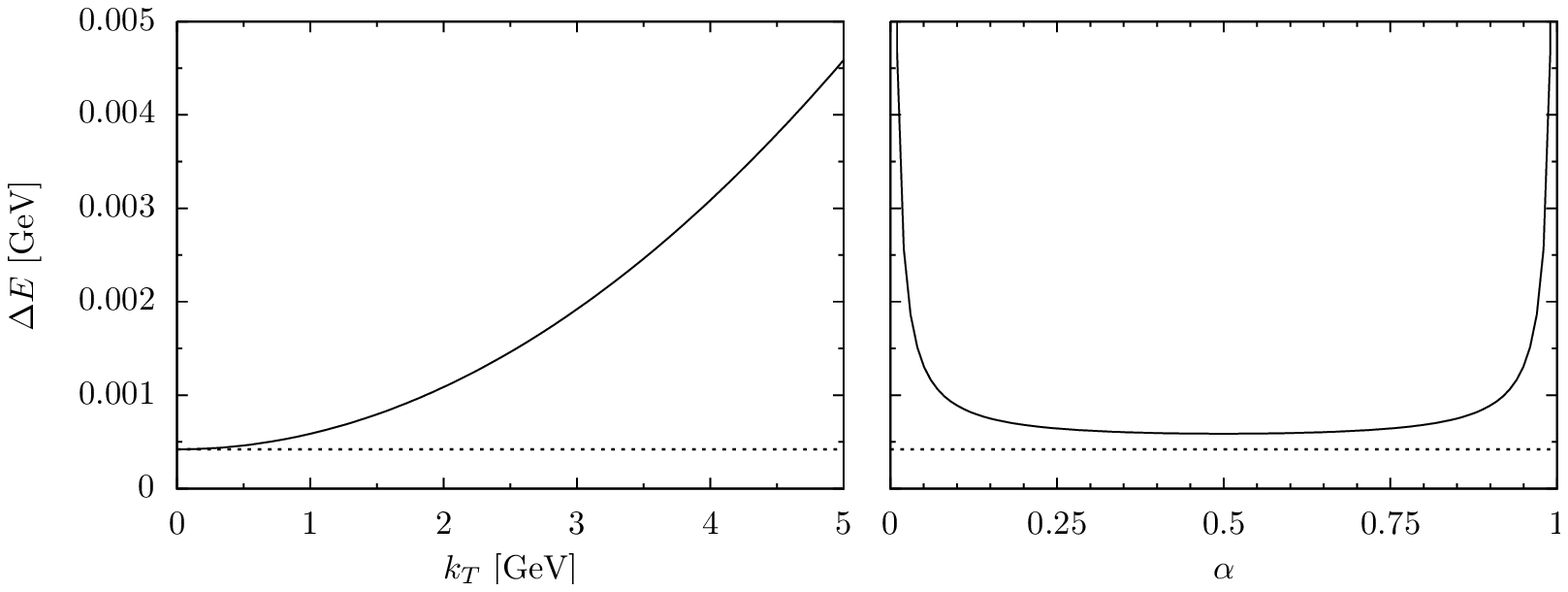}
\caption{Energy mismatch $\dE$ of the $\gamma^*$ dipole transition 
as a function of the absolute value $k_T$ of the quark's transverse momentum 
(left graph) and of its longitudinal momentum fraction $\alpha$ (right graph).
The parameters are $W=150$~GeV, $Q^2=10$~GeV$^2$, $m_q=0.14$~GeV.
For the left graph we have chosen $\alpha=1/2$ and for the right graph $\kt=1$~GeV, 
respectively. 
The dotted line shows the absolute minimum $\dE_{\text{min,abs}}$ for the
given external kinematics, that is, $\dE$ at $\kt=0$ and $\alpha=1/2$.
\label{fig:deltaektal}}
}
At fixed $\alpha$ the energy mismatch becomes minimal for $k_T=0$: 
\begin{equation}
\label{dEminal}
\dE_{\text{min}}(\alpha) 
= \dE(0,\alpha) = \sqrt{\alpha^2\nvec{q}^2+m_q^2}
   + \sqrt{(1-\alpha)^2\nvec{q}^2+m_q^2}
   - q^0\,.
\end{equation}
The absolute minimum of $\dE$ is reached at $k_T=0$, $\alpha=1/2$,
\begin{equation}
\label{dEmintot}
\dE_{\text{min,abs}} = \dE(0,1/2) = \sqrt{\nvec{q}^2 + 4 m_q^2}
   - q^0\,.
\end{equation}
That is, we have 
\begin{equation}
\dE(k_T,\alpha) \geq \dE_{\text{min}}(\alpha) \geq \dE_{\text{min,abs}} \geq 0\,,
\end{equation}
where the last inequality is strict for all $Q^2>0$.
Thus we see that there is an a priori minimal value for the energy mismatch
$\dE$ at a given kinematical point and for given quark flavour. 
Figure\ \ref{fig:deltaemin} in section \ref{sec:results} 
shows twice this minimal energy mismatch. 

\section{Some technical details of the calculation}
\label{appB}

In this appendix we give the details of the calculations for the $\Delta E$ 
distributions \eqref{xsdetadetab}, \eqref{xstldEp}, and the corresponding 
quantities for individual quark flavours $q$. 
We use the Golec-Biernat-W\"usthoff 
model \cite{GolecBiernat:1998js} which describes the $F_2$ data 
from HERA quite well. The dipole cross section of this model is given in 
\eqref{xsgbw} and \eqref{GBWparam}. We point out again that the 
dipole cross sections $\hat{\sigma}_{\mathrm{GBW}}^{(q)}$ 
depend not only on $W$ but also on $Q^2$ through the $x$ dependence. 
We use the GBW model nevertheless, since we shall only be concerned with 
specific kinematic values of $W$ and $Q^2$ for which we study Ioffe-time 
distributions. We shall not compare structure functions at the same 
$W$ and different $Q^2$ values, where the choice of energy variable 
is essential, see \cite{Ewerz:2006an,Ewerz:2007md}. 

The Fourier transformation \eqref{xsdipfourier} 
of $\hat{\sigma}_{\mathrm{GBW}}^{(q)}(r,x)$
\eqref{xsgbw} gives
\begin{equation}
\label{fouriergbw}
\tilde{\hat{\sigma}}_{\mathrm{GBW}}^{(q)}(\vkt,x)
  = \sigma_0 \left[ (2\pi)^2 \delta^{(2)}(\vkt)
                   - 4 \pi r_0^2  \exp(-r_0^2 \kt^2) \right] \,.
\end{equation}
Our aim is to calculate the $\Delta E_+$ distributions for 
the various quark-flavour contributions to $\sigma_T$ and $\sigma_L$. 
We have 
\begin{equation}
\label{xsdetadetab-mod}
\begin{split}
\frac{\partial^2 \sigma_{T,L}^{(q)} (W,Q^2,\eta,\bar{\eta})}{\partial\etaa\,\partial\etab} =
  \sum_{\lambda,\lambda'} \int\! \ud\alpha &
  \int\! \frac{\ud^2 \kbt}{(2\pi)^2}
  \int\!  \frac{\ud^2 \kat}{(2\pi)^2}
  \left( \tilde{\psi}^{(q)\,\pm,L}_{\lambda\lambda'}(\alpha,\vkbt,Q)
  \right)^\ast\,
  \delta(\etab-\dEb)
\\ 
&\times   \tilde{\sigmadip}^{(q)}(\vkat-\vkbt,W)\,
  \delta(\etaa-\dEa)\,
  \tilde{\psi}^{(q)\,\pm,L}_{\lambda\lambda'}(\alpha,\vkat,Q)\,,
\end{split}
\end{equation}
\begin{align}
\label{xstldEp-q}
\frac{\partial\sigma_{T,L}^{(q)}(W,Q^2,\dEp)}{\partial\dEp}
&= \int_0^\infty\! \ud\eta \int_0^\infty\! \ud \bar{\eta}\,
  \frac{\partial^2 \sigma_{T,L}(W,Q^2,\eta,\bar{\eta})}{\partial\eta\,\partial\bar{\eta}}\,
  \delta(\dEp-\eta-\bar{\eta}) \,.
\end{align}
We insert the photon wave functions from 
\eqref{psitkgen} and \eqref{psilkgen} 
and replace $\tilde{\hat{\sigma}}^{(q)} (\vkt - \vkbt,W)$ by 
$\hat{\sigma}_{\text{GBW}}^{(q)} (\vkt- \vkbt,x)$ 
from \eqref{fouriergbw}. 
In order to integrate out the azimuthal angles we decompose
the photon wave functions as follows 
\begin{equation}
\tilde{\psi}^{(q)\,\pm,L}_{\lambda\lambda'}(\alpha,\vkat,Q) =
  \sum_{n=-2}^{2} e^{i n \phi_k}
    \tilde{\psi}^{(q)\,\pm,L}_{\lambda\lambda' n}(\alpha,\kat,Q) \,.
\end{equation}
For any function $f(k_T)$ we have the relation 
\begin{equation}
\int_0^\infty \ud\kt\, f(\kt) =
  \int_{\dE_{\text{min}}(\alpha)}^\infty \ud(\dE)\,
  \frac{k^0 k'^0}{\kt(k^0+k'^0)} f(\kt(\dE,\alpha)) \,.
\end{equation}
Using this we find for the distributions in $\dEp$ that 
\begin{equation}
\label{dsigdEpgbw}
\frac{\partial\sigma_{T,L}^{(q)}}{\partial\dEp} = \,
  \frac{\partial\sigma_{T,L}^{(q),\text{const}}}{\partial\dEp}
 +\frac{\partial\sigma_{T,L}^{(q),\text{e}}}{\partial\dEp}
\end{equation}
with
\begin{align}
\label{dsigdEpgbwp1}
\frac{\partial\sigma_{T,L}^{(q),\text{const}}}{\partial\dEp} =& \,
  \sigma_0
  \int_{\alpha_{\text{min}}(\dEp)}^{\alpha_{\text{max}}(\dEp)}\hspace{-9ex}
  \ud\alpha\hspace{6ex}\,
  \frac{\ka^0\ka'^0}{2\pi(\ka^0+\ka'^0)}
  \sum_{\lambda,\lambda',n}
  \abs{\tilde{\psi}^{(q)\,\pm,L}_{\lambda\lambda' n}(\alpha,\kat,Q)}^2\,,\\
\label{dsigdEpgbwp2}
\frac{\partial\sigma_{T,L}^{(q),\text{e}}}{\partial\dEp} =& \,
  \sigma_0
  \int_{\alpha_{\text{min}}(\dEp)}^{\alpha_{\text{max}}(\dEp)}\hspace{-9ex}\ud\alpha\hspace{6ex}
  \int_{-\dE_{\text{min}}(\alpha)}^{\dE_{\text{min}}(\alpha)}\hspace{-8ex}\ud\dEm\hspace{3ex}\,
  \frac{\ka^0\ka'^0\kb^0\kb'^0 r_0^2}{2\pi(\ka^0+\ka'^0)(\kb^0+\kb'^0)}
e^{-r_0^2(\kat-\kbt)^2} 
 \notag\\
	& 
  \times 
  \sum_{\lambda,\lambda',n} \frac{I_n(2 r_0^2\kat\kbt)}{\exp(2 r_0^2\kat\kbt)}\,
  \tilde{\psi}^{(q)\,\pm,L}_{\lambda\lambda' n}(\alpha,\kat,Q)\,
  \left(\tilde{\psi}^{(q)\,\pm,L}_{\lambda\lambda' n}(\alpha,\kbt, Q)\right)^\ast\,,
\end{align}
where $I_n$ are the modified Bessel functions. 
Here, the energies and momenta of quark and anti-quark \eqref{kquarkdecomp}
are understood as functions of $\alpha$, $\dEa$, $\dEb$ via
\begin{align}
\kat &= \sqrt{\frac{\left( (\dEa+q^0)^2 - \nvec{q}^2 \right)
                  \left( (\dEa+q^0)^2 - (1-2\alpha)^2 \nvec{q}^2 \right)}
                 { 4 (\dEa+q^0)^2 }
            -m_q^2 }\,,
\\
\kbt &= \sqrt{\frac{\left( (\dEb+q^0)^2 - \nvec{q}^2 \right)
                  \left( (\dEb+q^0)^2 - (1-2\alpha)^2 \nvec{q}^2 \right)}
                 { 4 (\dEb+q^0)^2 }
            -m_q^2 }\,.
\end{align}
The $\alpha$-integration range in \eqref{dsigdEpgbwp1} is finite for fixed $\dEp$ 
and its endpoints correspond to vanishing transverse momenta.
The extremal values of $\alpha$ are given by the two solutions of the equation
$\dEp/2 = \left.\dE\right\vert_{\kt=0}$ with respect to $\alpha$:
\begin{equation}
\label{alphaminmax}
\alpha_{\text{max}, \text{min}} 
= \frac{1}{2}
  \pm \frac{\dEp/2+q^0}{2\abs{\nvec{q}}}
      \sqrt{1-\frac{4 m_q^2}{(\dEp/2+q^0)^2-\nvec{q}^2}}\,.
\end{equation}
In all cases considered in this paper $\alpha_{\text{min}}\approx 0$ and
$\alpha_{\text{max}}\approx 1$.

We perform the residual integrations in \eqref{dsigdEpgbwp1} and 
\eqref{dsigdEpgbwp2} numerically.
We avoid numerical errors due to the large cancellations in sums 
like \eqref{dEktal} by proper rewriting, and we make sure that the integration 
error is under control by using different
integration algorithms \cite{Hahn:2004fe,gsl18}.

\end{appendix}

\end{document}